\documentclass[aps,prd,twocolumn,groupedaddress,nofootinbib]{revtex4-1}
\usepackage{graphicx,color,amsmath}
\usepackage[export]{adjustbox}
\usepackage{slashed}
\usepackage{booktabs} 
\usepackage{braket}
\usepackage{hyperref}
\usepackage{mathtools} 

\usepackage{tikz}
\usepackage{tkz-euclide}
\usetikzlibrary{decorations.text}
\usetikzlibrary{decorations.pathmorphing}	
\usetikzlibrary{fadings}
\tikzset{
    v/.style={decorate, decoration={snake, segment length=3mm, amplitude=0.75mm}, draw},
    f/.style={draw,decoration={markings,mark=at position #1 with {\arrow[very thick]{latex}}},postaction={decorate},node contents=#1},
    f/.default=.6,
    fb/.style={draw,decoration={markings,mark=at position #1 with {\arrowreversed[very thick]{latex}}},postaction={decorate},node contents=#1},
    fb/.default=.4,
    fnar/.style={draw},
    g/.style={decorate, draw,  decoration={coil,amplitude=3pt, segment length=3.5pt}},
    s/.style={dashed,draw, postaction={decorate},
        decoration={markings,mark=at position .55 with {\arrow[very thick]{latex}}}},
    sb/.style={dashed,draw, postaction={decorate},
        decoration={markings,mark=at position .55 with {\arrowreversed[draw=black,very thick]{latex}}}},
    snar/.style={dashed,draw,line width =1.25pt},
}

\definecolor{cblue}{rgb}{0., 0., 1.}
\definecolor{cgreen}{rgb}{0.,0.666667, 0.}
\definecolor{cpink}{rgb}{0.9, 0.09, 0.72}
\definecolor{cred}{rgb}{.8,0,0}
\definecolor{corange}{rgb}{1., 0.5, 0.}
\definecolor{cgray}{rgb}{0.5, 0.5, 0.5}

\begin{document}

\title{Discovering leptonic forces using non-conserved currents}

\author{Jeff A. Dror}
\email{jdror@lbl.gov} 
\affiliation{Theory Group, Lawrence Berkeley National Laboratory, Berkeley, CA 94720, USA}
\affiliation{Berkeley Center for Theoretical Physics, University of California, Berkeley, CA 94720, USA}

\date{\today}


\begin{abstract}
Differences in lepton number (i.e., $ L _e - L _\mu $, $ L _e - L _\tau $, $ L_\mu  - L _\tau  $, or combinations thereof) are not conserved charges in the Standard Model due to the observation of neutrino oscillations. We compute the divergence of the corresponding currents in the case of Majorana or Dirac-type neutrinos and show that, in the high energy limit, the vector interactions map onto those of a light scalar coupled to neutrinos with its coupling fixed by the observed neutrino masses and mixing. This leads to amplitudes with external light vectors that scale inversely with the vector mass. By studying these processes, we set new constraints on $ L _i - L _j $ through a combination of semi-leptonic meson decays, invisible neutrino decays, neutrinoless double beta decays, and observations of Big Bang Nucleosynthesis/supernova, which can be much stronger than previous limits for vector masses below an eV. These bounds have important implications on the experimental prospects of detecting $ L _i - L _j $ long-range forces.
\end{abstract}

\maketitle

\section{Introduction}
Light vector bosons ($ X ^\mu $) are simple, technically natural, extensions to the Standard Model of particle physics, which can play a key role in explaining experimental anomalies~\cite{gninenko:2001hx,kahn:2007ru,pospelov:2008zw,tuckersmith:2010ra,batell:2011qq,feng:2016ysn,altmannshofer:2017bsz}, act as a mediator to the dark sector~\cite{boehm:2003hm,pospelov:2007mp,arkanihamed:2008qn}, or make up dark matter themselves~\cite{Arias:2012az,Graham:2015rva,Dror:2018pdh,Bastero-Gil:2018uel,Co:2018lka,Long:2019lwl}. If a vector mass ($ m _X $) is well below the weak scale then, depending on the type of current involved, experimental signals have drastically different behavior as a function of $ m _X $. For a vector coupled to the electromagnetic current (i.e., a kinetically mixed dark photon), then as $ m _X \rightarrow 0 $ it becomes an unobservable correction to electromagnetism resulting in all constraints deteriorating at low masses. On the other hand, if it couples to any other conserved current, $  j _\mu $, (i.e., a current which satisfies $ \partial _\mu j  ^\mu = 0 $ at the quantum level) then the constraints become mass-independent as $ m _X \rightarrow 0 $. Lastly, if a vector is coupled to a non-conserved current, there exist processes whose amplitudes are proportional to $ 1 / m _X $, leading to constraints that get stronger at low masses (see ~\cite{Fayet:2006sp,Barger:2011mt,Laha:2013xua,Karshenboim:2014tka,Dror:2017nsg,Dror:2017ehi,Dror:2018wfl,Arcadi:2018xdd} for cases where this property was previously used to set bounds on light vectors). 

The Standard Model with massless neutrinos has a set of four linearly-independent conserved currents:~\footnote{$ j _{ L _e -  L _\tau } $ is a linear combination of $ j _{ L _e - L _\tau } $ and $ j _{ L_\mu - L _\tau} $.}
\begin{equation} 
  j _{ {\rm EM}}, \quad j _{ B - L} , \quad j _{ L _e - L _\mu } , \quad j _{ L _\mu  - L _\tau } \,.
\end{equation} 
where $ j _{ {\rm EM}} $ is the electromagnetic current, $  B $ and $ L $ are baryon and lepton number respectively, and $ L _i $ denotes lepton number of generation, $ i $. Once neutrino masses are introduced the only currents which remain conserved are electromagnetism and $ B - L $ (assuming neutrinos are of Dirac-type); The ``flavored'' leptonic currents, $ j_{L _i - L _j} $, cannot explain the observed neutrino masses and mixing without introducing a small breaking, typically assumed to be from integrating out some scalar(s) and right-handed neutrinos~\cite{Ma:2001md,Choubey:2004hn,Heeck:2011wj,Asai:2018ocx}. \footnote{These mechanisms are generically easier to build when $ m _X / g _X \gtrsim 100 ~{\rm GeV} $ since then the vacuum expectation values of any U(1)$ _{ L _i - L _j } $-breaking scalar can contribute evenly to all the neutrinos, thereby preventing any of the lepton mixing angles from being too small. However, we do not take this as a bound as one can always evade this theoretical challenge by introducing multiple scalar fields.} While the inevitable breaking has negligible impact when $ m _X \gtrsim  ~{\rm eV}$, it can have dramatic effects at lower masses, in the form of enhanced production of the vector's longitudinal mode. Our goal is to explore these effects. 

Since the symmetry is broken by the observed neutrino masses and mixing it will lead to amplitudes proportional to $ ( g _X m _\nu / m _X ) $, and hence most prominent at low $ m _X $. As we will show, a combination of constraints from semi-leptonic meson decays ($ K ^\pm \rightarrow \ell  ^\pm \nu X $, $ \pi ^\pm \rightarrow \ell  ^\pm \nu X $), lepton-flavor violating processes ($ \nu _i \rightarrow \nu _j X $, $ X $-emitting neutrinoless double beta decays), and neutrino annihilations (changing $ \Delta N _{ {\rm eff}} $/supernova), result in important bounds on the $ L _i - L _j $ parameter space. To keep the discussion general we do not specify a particular mass mechanism for the neutrinos. While this has the drawback that processes may develop sensitivity to the UV, most do not, and we focus on bounds that are robust. 

Aside from the constraints we present here, the best constraints on gauged $ L _i - L _j $ depend strongly on whether the difference in lepton number includes electron number. For vectors coupled to $ j_{L _e - L _\mu} $ and $ j_{L _e - L _\tau} $ with $ m _X \lesssim {\cal O} ( 0.1 ~{\rm eV} ) $ the most stringent current limits are from fifth force/EP-violation searches~\cite{Wise:2018rnb,Kapner:2006si,Schlamminger:2007ht,Salumbides:2013dua} and modifications to neutrino oscillations due to a long-range potential induced by the Earth, Sun, or whole galaxies~\cite{Joshipura:2003jh,Grifols:2003gy,Bandyopadhyay:2006uh,GonzalezGarcia:2006vp,Samanta:2010zh,Wise:2018rnb,Bustamante:2018mzu}, with the gauge coupling ($ g _X $) being constrained to be weaker than gravity for $ m _X \lesssim {\cal O} (  10 ^{ - 3} {\rm eV} )   $. 

For $ L _\mu - L _\tau $, where common matter does not induce a long-ranged force, the dominant low mass constraints are from dynamics of neutron star inspirals~\cite{Dror:2019uea} as well as from pulsar binaries \cite{Poddar:2019wvu,Dror:2019uea}. For $ m _X \gtrsim 10 ^{ - 11} ~{\rm eV} $ (the inverse size of a neutron star) these bounds do not apply and the most stringent bounds are from (non-enhanced) neutrino annihilations to vectors increasing $ \Delta N _{ {\rm eff}} $. If the universe reheated above the muon mass after inflation and there were no additional entropy dumps after muons froze-out, then the best constraints are from muon annihilations to vectors $ \mu ^+ \mu \rightarrow \gamma X $ constraining $ g _X \lesssim 4 \times 10 ^{-9 } $~\cite{Escudero:2019gzq} (see~\cite{Grifols:1996gn,Dolgov:1999gk} for earlier work). Otherwise, the strongest limits are due to neutrino annihilations constraining $ g _X \lesssim 5 \times 10 ^{ - 6} $~\cite{Huang:2017egl}. We will consider the same process but focusing on the enhanced contributions in section~\ref{sec:ann}. For light bosons there are additional constraints from the observation of near-extremal black holes and the null observation of superradiance~\cite{Arvanitaki:2009fg,Baryakhtar:2017ngi}. While we include these, they could change dramatically if the scalar field used to provide to break U(1)$ _{ L _i - L _j } $ is light enough to induce self-interactions within the superradiance cloud~\cite{Baryakhtar:2017ngi}. 

The constraints presented here, dramatically shape the low mass $ L _i - L _j $ parameter space. In particular, for $ L _e - L _\mu $ and $ L _e - L _\tau $ the enhanced processes are the strongest bounds for $ m _X \lesssim 10 ^{ - 22 }  - 10 ^{ - 17}~{\rm eV} $, depending the theory assumptions. For $ L _\mu - L _\tau $ the new constraints can be the strongest for all $ m _X \lesssim 100~ {\rm eV}  $ or for $ {\rm eV} \lesssim m _X \lesssim 10 ^{ - 10}~ {\rm eV}  $ and $ m _X \lesssim 10 ^{ - 15} ~{\rm eV} $, depending on the theory assumptions. 

The paper is organized as follows. In section~\ref{sec:nonconserved} we compute the divergence of $ L _i  - L _j $ current and study the correspondence between $ L _i  - L _j $ vectors and light scalars. We use these to compute constraints from semi-leptonic meson decays in section~\ref{sec:meson}, neutrino decays in section~\ref{sec:nudecay}, neutrinoless double beta decays in section~\ref{sec:0nu2b}, and neutrino annihilations prior to Big Bang Nucleosynthesis and inside supernova in section~\ref{sec:ann}. We conclude by discussing the implications of these new bounds focusing on the prospects of seeing deviations to neutrino oscillations from long-range forces in section~\ref{sec:disc}.
\section{$ L _i - L _j $ and non-conserved currents}
\label{sec:nonconserved}
We begin by summarizing gauged $ L _i - L _j $ models, with an emphasis on the parts relevant for non-conserved currents. The coupling of $ X ^\mu $ to the Standard Model fermions is given in terms of two-component Weyl spinors by, \footnote{We follow the conventions of two-component spinors laid out in~\cite{Dreiner:2008tw}.}
\begin{align} 
& {\cal L} _X  = g  _X X ^\mu \hat{Q}  _{ii}\left(  \hat{ \ell} _i ^\dagger \bar{\sigma}  _\mu \hat{ \ell} _i - \hat{ \ell} _i ^{c \dagger} \bar{\sigma}  _\mu \hat{ \ell} _i ^c  +  \hat{ \nu} _i ^\dagger \bar{\sigma} _\mu \hat{ \nu}_i   \right) \,, 
\end{align}
where we denote right-handed fields using a $ {} ^c $ and write all quantities in the flavor basis with $ \hat{{}}$~.  The U(1)$_{ L _i -  L _j }$ charge matrix, $\hat{Q}  _{ ij} $ depends on the particular flavor symmetry and is $  = {\rm diag} (1,-1,0) $, ${\rm diag} (1,0,-1) $, or ${\rm diag} (0,1,-1) $ for $  L _e - L _\mu $, $ L _e - L _\tau $, or $ L _\mu - L _\tau $ respectively. In addition, there may be couplings of right-handed neutrinos to $ X $, $ X ^\mu \hat{Q}  ^c  _{ii} \hat{ \nu} _i ^{ c \dagger } \bar{\sigma} _\mu \hat{ \nu} ^c _i $, with an {\em a priori} unknown charge matrix, $ \hat{Q}  _i ^c $, which only needs to satisfy the anomaly cancellation condition, $ {\rm Tr} \hat{Q} ^{c\,3} = 0 $. We discuss the implications of the right-handed neutrino-$ X $ interaction in some detail below.  

Imposing one of the lepton flavor symmetries, while assuming the U(1)$ _X $-breaking sector does not contribute to the lepton masses, prevents mixing in the charged lepton sector (i.e., $ \hat{ \ell } _i = \ell _i $) and hence the observed lepton mixing which makes up the Pontecorvo–Maki–Nakagawa–Sakata (PMNS) matrix, $ U $, arises entirely from the neutrino mixing matrix (i.e., $ \hat{\nu}  _i = U _{ ij} \nu _j $). As such, performing a rotation to the mass basis,
\begin{align} 
{\cal L} _X  & =  g  _X X ^\mu \left(   \hat{Q}  _{ii} \ell _i ^\dagger \bar{\sigma}  _\mu  \ell _i -  \hat{Q} _{ii} \ell _i ^{c \dagger} \bar{\sigma}  _\mu  \ell _i ^c  + ( U ^\dagger \hat{Q} U ) _{ ij}\nu _i ^\dagger \bar{\sigma} _\mu \nu _j   \right)\,,\label{eq:lagX}
\end{align} 
changes the form of the coupling to neutrinos but leaves the charged lepton interaction diagonal. The neutrino mass term depends on whether neutrinos are Majorana (M) or Dirac (D),~\footnote{The case of pseudo-Dirac neutrinos will be somewhere in-between the two possibilities studied here.}
\begin{equation} 
\begingroup
\renewcommand*{\arraystretch}{1.25}
- {\cal L} _{ {\rm mass}} =  \left\{ \begin{array} {lc}\frac{1}{2} \hat{m}  _{ ij}   \hat{ \nu} _i \hat{ \nu} _j  & ({\rm M}) \\ \hat{m}  _{ij}  \hat{ \nu} ^c _i \hat{ \nu} _j    & ({\rm D})  \end{array} \right.  +{\rm h.c.} \,,
\endgroup
\end{equation} 
In either case imposing $ L _i - L _j $ cannot reproduce the PMNS matrix. This is simple to see if neutrinos are Majorana since $ L _i - L _j $ would restrict $ \hat{m}  _{ij}  $ from having any mixing between one of the flavors from the other two and hence cannot possibly reproduce nonzero values for all three $ 3 $ mixing angles. 

If neutrinos are Dirac then the situation is more subtle. Upon diagonalization the mass term is given by,
\begin{equation} 
- {\cal L} _{ {\rm mass}} =  ( V ^\dagger m  U ) _{ ij } \hat{\nu}  _i ^c \hat{\nu}  _j +{\rm h.c.} \,,
\end{equation} 
where $ V $ is the right-hand neutrino rotation matrix necessary to diagonalize $ \hat{m}  _{ ij} $ such that $ V ^\dagger   \hat{m} U   = m \equiv {\rm diag} ( m _{ 1} , m _{ 2 } , m _3 ) $. $ V $ is undetermined until a choice is made for the charges of the right-handed neutrinos, $ \hat{Q}  ^c  _i $. However, even allowing for arbitrary right-handed neutrino charges, the observed masses and mixing violate $ L _i - L _j $ symmetry. To see this explicitly consider the effect of a $ L _i - L _j $ rotation on the neutrino mass masses. In order for the term to be $ L _i - L _j $ invariant we require 
\begin{align} 
 \hat{Q}  ^{c\, T}  V ^\dagger m  U \hat{Q}   & = V ^\dagger m  U  \,,\\ 
\Rightarrow   ( V \hat{Q}  ^{c\, T}  V ^\dagger )   m  U \hat{Q}   & = m  U \,.
\end{align} 
For an arbitrary matrix, $ V \hat{Q}  ^{c\,T}  V ^\dagger $, this equation has no solution since the form of $ \hat{Q}   $ requires the left-hand side to have a vanishing column while $ m  U $ has been measured to be entirely non-vanishing. We conclude that the observed neutrino masses and their mixing always break U(1)$ _{ L _i - L _j } $.

The consequence of this breaking can be seen by computing the divergence of the current and we carry this out in detail in appendix~\ref{app:replacement}. The result is,
\begin{equation} 
\begingroup
\renewcommand*{\arraystretch}{1.25}
\partial _\mu j _{ L _i - L _j } ^\mu =  i \left\{ \begin{array}{lc}    \hat{Q}  _{ a } \hat{m}  _{ab} \hat{\nu}  _a \hat{\nu}  _b &   ({\rm M})  \\   \left[ \hat{Q}   _{ b } \hat{m}  _{ab} + \hat{Q}  ^c  _{ a } \hat{m}  _{ab} \right]  \hat{\nu}  _a ^c \hat{\nu}  _b  & ({\rm D}) \end{array} \right.  +{\rm h.c.}\,,
\endgroup
\label{eq:div1}
\end{equation} 
or in terms of the mass eigenstates,
\begin{equation} 
\begingroup
\renewcommand*{\arraystretch}{1.25}
\partial _\mu j _{ L _i - L _j } ^\mu =   i \left\{ \begin{array}{lc}  Q _{ ab } m _{b} \nu _a \nu _b & ( {\rm M} )   \\   \left[ Q  _{ a b } m _a + Q ^c  _{ ab } m _b \right]  \nu _a ^c \nu _b  & ( {\rm D} )  \end{array}   \right. +{\rm h.c.}\,,
\endgroup
\label{eq:div2}
\end{equation} 
where we have introduced some convenient notation to eliminate extra $ U $s when possible: $ Q \equiv U ^T \hat{Q} U $ (Majorana) or $ Q \equiv U ^\dagger \hat{Q} U $ (Dirac) and $ m _{ 1,2,3 } $ are the neutrino masses in ascending order.~\footnote{When explicit numbers are necessary, we will take the best fit numbers from \cite{Esteban:2018azc} for a normal ordered hierarchy such that $ m _3 \simeq \sqrt{ \Delta m _{ {\rm atm}} ^2 } \simeq 0.05 {\rm eV} $, $ m _2 \simeq \sqrt{ \Delta m _{ {\rm sol}} ^2 } \simeq 0.087 {\rm eV} $, $ m _1 \simeq 0$ though none of our observables are particularly sensitive to $ m _1 $ or the neutrino hierarchy.}We will use $ Q $ or $ \hat{Q} $ depending on which is more convenient.

In the Majorana case, the divergence is manifestly nonzero if $ Q _{ij}  $ and $ m _{ i}  $ are nonzero, while for the Dirac case one can show that there is no choice of $ Q _{ij} ^c  $ that allows the Dirac term to vanish, which is simply a restatement of the above observation that the presence of neutrino masses and mixing inevitably breaks the $ L _i -  L _j $ symmetry. For all processes studied in this work, a right-handed neutrino coupling to $ X $ would only add to the rates at leading order in the neutrino mass (which will be an expansion parameter). To be conservative we simply set $ Q ^c = 0 $ throughout but note that if the right-handed neutrinos are charged under U(1)$ _{ L _i - L _j } $, it would strengthen our prospective bounds. 

The impact of this term becomes clear employing Goldstone boson equivalence theorem (GBET), which states that, in the high energy limit (energy of $ X $ much larger than its mass), we can compute amplitudes of $ X ^\mu $ employing the replacement, $ X ^\mu \rightarrow   g _X i \partial ^\mu \varphi _X  / m _X $, and computing the amplitudes for a scalar $ \varphi _X  $ instead (see appendix \ref{app:replacement} for more details). The $ \varphi _X  $ interactions, in the mass basis, are then governed by,
\begin{align} 
{\cal L} _X & = X _\mu j _{ L _i - L _j } ^\mu \,, \notag \\ 
& \begingroup
\renewcommand*{\arraystretch}{1.25}
\rightarrow \frac{ i g _X Q _{ ab } m _a }{ m _X }  \varphi _X  \left\{ \begin{array}{lc}        \nu _a \nu _b   & ({\rm M}) \\      \nu _a ^c \nu _b  & ({\rm D})   \end{array}   \right. +{\rm h.c.}\,.\endgroup
\label{eq:GBET}
\end{align} 
To obtain this expression we have integrated by parts to replace $ \partial _\mu j ^\mu _{ L _i - L _j } $ using \eqref{eq:div1}, setting $ Q ^c \rightarrow 0 $. For boosted external $ X ^\mu $, this results in amplitudes proportional to $ g  _X m _\nu  / m _X $. These will always be parametrically the most important processes as $ m _X \rightarrow 0 $. We see that, at high energies $ L _i - L _j $ vectors are effectively Majorons~\cite{Gelmini:1980re} for Majorana neutrinos and neutrino-philic scalars for Dirac neutrinos. Of course one is always free to either use the Goldstone boson expression, \eqref{eq:GBET} or the vector Lagrangian, \eqref{eq:lagX} (though this correspondence breaks down when $X ^\mu $ is virtual). 

Any UV complete model will contribute new sources to the right-hand side of \ref{eq:div1} or \ref{eq:div2} through the field(s) that spontaneously break the symmetry and adding these could change our results. In particular since we can be in the regime where $ m _X / g _X \sim {\rm eV} $, it is likely other degrees of freedom can be produced on-shell in processes closely related to the ones studied here. While this would add to the rates, these are more model-dependent and we do not include them. Furthermore, using these additional degrees of freedom to eliminate the processes computed here would likely introduce an enormous fine-tuning and would be challenging given the multitude of possible signals. 

Lastly, we comment on the minimal requirement for theories with a non-conserved current - ensuring unitarity; If $ g _X / m _X $ is too large, then in any diagram with neutrinos, we can add an additional external $ X $ and increase the amplitude, leading to a breakdown of unitarity. Since a $ g _X / m _X $-enhancement will always cost a power of a neutrino mass, the unitarity condition is,
\begin{equation} 
\frac{ g _X }{ m _X } \lesssim \frac{ 4\pi  }{ m _\nu } \,.
\end{equation} 
The factor of $ 4\pi $ arises from the phase space cost of adding in an extra particle. The unitarity bound applies equally to both Majorana and Dirac neutrinos and is robust against the details of the UV completion (up to $ {\cal O} ( 1 ) $ corrections). We will impose this throughout as an orange region in our constraint plots.

\section{Meson decays ($ P ^\pm  \rightarrow \ell _b ^\pm   \nu _a X $)} 
\begin{figure*} 
\begin{center} 
  \includegraphics[width=13.5cm]{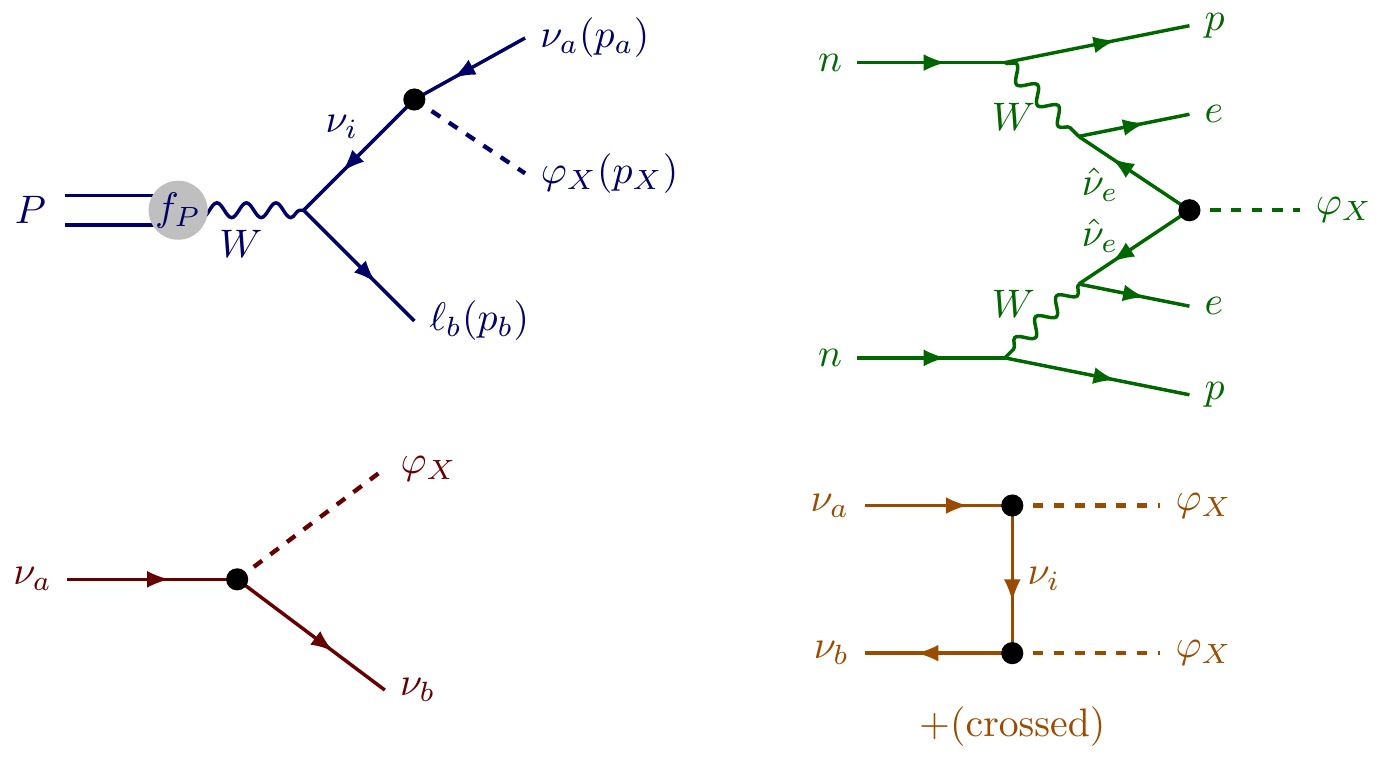}
\end{center}
\caption{Different processes used to constrain $  L _i - L _j $ vectors with diagrams shown as computed in the Goldstone boson equivalent theories. We show  semi-leptonic meson decay ({\bf \color{blue!40!black} top-left}), neutrino decay ({\bf \color{red!40!black}bottom-left}), neutrinoless double beta decays ({\bf \color{green!40!black}top-right}), and neutrino annihilations ({\bf \color{orange!60!black}bottom-right}). Black dots indicate the $ L _i - L _j $ Goldstone boson emission point. For neutrinoless double beta decay, the internal neutrinos are in the flavor basis as emphasized by the $ \hat{ {}} $~.}
\label{fig:allprocesses}
\end{figure*}
\label{sec:meson}
As our first enhanced process we study semi-leptonic meson decays, $ P ^\pm  \rightarrow \ell _b ^\pm  \nu _a X $ ($ P = B , D, K, \pi $). This process is depicted in the Goldstone picture in Fig.~\ref{fig:allprocesses} ({\bf top-left}), where the scalar is radiated off the neutrino leg. In the vector picture there are two diagrams that one must add together, and a delicate cancellation takes place for the longitudinal mode leaving only the piece proportional to $ m _\nu g _X / m _X $. This process was previously considered in~\cite{Ibe:2016dir} to probe $ L _i - L _j $ however they did not include the important enhanced contribution for small $ m _X $. Energy-enhanced semi-leptonic decays to $ \ell \nu X $ were discussed for parity violating muon couplings~\cite{Barger:2011mt} and for vectors coupled only to neutrinos~\cite{Laha:2013xua,Bakhti:2017jhm,Bahraminasr:2020ssz}. Such vectors are coupled to currents whose divergence is proportional to the lepton masses and hence have amplitudes that grow as $ \propto \,m _\ell / m _X $.

The decay rate is different in the case of Majorana and Dirac neutrinos and we compute both. The final result is,
\begin{equation} 
\Gamma _{ P ^- \rightarrow \ell _b  ^- \nu _a  X} =  \frac{M _P }{ 256 \pi ^3  } \int _0 ^{1 - \alpha ^2 } d x _1 \int _{ 1 - \alpha ^2 - x _1} ^{1 - \alpha ^2 / ( 1 - x _1 ) }  d x _2 \overline{ \left| {\cal M} \right| ^2 }\,,
\end{equation} 
where phase space integral is given in term of the energy fractions $ x _{ 1 }  \equiv 2 E _X / M _P $, $ x _2 \equiv 2 E _{\nu _a } / M _P $, and we defined $ \alpha \equiv M _{\ell } / M _P $, the ratio of the lepton to meson mass, and neglected the $ X ^\mu $ and external neutrino masses. The spin-averaged square of the matrix element for Majorana (top line) and Dirac (bottom line) neutrinos is,
\begin{align} 
  \overline{ \left| {\cal M} \right| ^2 } & = \left| \frac{ g ^2 f _P}{ 2 m _W ^2 }  V _{ {\rm CKM}} \frac{ g _X }{ m _X }\right| ^2 M _P ^2 \,I (x _1 , x _2 )   \,,\notag \\ 
& \qquad \times \left\{ \begin{array}{l} \left| \sum _i Q _{a i} U _{ b i } ( m _a + m _i )  \right| ^2   \\  \left|   \hat{Q} _{ b b}U _{ b a }   m _a  \right| ^2 \end{array} \right.\,.
\end{align} 
$ V _{ {\rm CKM}} $ is the CKM matrix element on the hadronic side ( $ = 1 $ for pion decay, the sine of the Cabibo angle for Kaon decay, etc.), and $ f _{ P ^+ } $ is the meson factor ($ f _\pi \simeq 130.2 \pm 0.8 ~{\rm MeV} $~\cite{PhysRevD.98.030001}, $ f _K \simeq 155.7 \pm 0.3 ~{\rm MeV} $~\cite{PhysRevD.98.030001}, etc.), $ g \simeq 0.6 $ is the weak coupling, $ m _W $ is the $ W $ boson mass. The dimensionless function, $ I ( x _1 , x _2 ) $, is given by,
\begin{align} 
I ( x _1 , x _2 )  & =  \frac{ 2 P \cdot p _b P \cdot k  - M _P ^2  p _b \cdot k  }{ k \cdot p _a M _P ^2 }\,,\\ 
& = \frac{ 1 + x _1 ^2 + x _1 x _2 - 2 x _1 - x _2   - \alpha ^2 }{ 1 - x _1 - x _2 - \alpha ^2 }\,,
\end{align} 
where the momenta are defined in Fig.~\ref{fig:allprocesses} ({\bf top-left}). The phase space integral has an IR divergence when the internal neutrino goes on-shell, a feature emphasized and treated carefully in~\cite{Pasquini:2015fjv}. This arises from loop corrections of $ X ^\mu $ to the neutrino mass. The sensitivity to this divergence is mild and can be regulated by introducing a small mass for the internal neutrino and we employ this strategy here.

There are many experiments that looked for semi-leptonic charged meson decays. Due to the number of events, the best constraints arise from pion and kaon decays and we focus on those. Decays to electrons have a significant advantage as an additional vector can lift the usual 2-body helicity suppression, reducing the background rates. For rare pion decays a dedicated search for $ \pi ^\pm \rightarrow e ^\pm   \nu (X \rightarrow {\rm inv}) $ was performed in \cite{PhysRevD.37.1131} constraining the branching ratio to $ \lesssim 4 \times 10 ^{ -6} $, which we approximate as our limit. In addition, we note that the recent PIENU experiment has about $ \sim 10 ^3 $ times as many pions and could likely significantly improve this limit as was recently done for heavy neutrino searches in pion decays~\cite{Aguilar-Arevalo:2017vlf,Aguilar-Arevalo:2019owf}. No analogous search exists for $ \pi ^\pm \rightarrow \mu ^\pm \nu X $ though, given its additional experimental challenges, it is not likely to significantly improve the limits. 

There are no dedicated searches for $ K ^\pm \rightarrow e ^\pm \nu X $ or $ K ^\pm \rightarrow \mu ^\pm \nu X $ however E949 looked for $ K ^\pm  \rightarrow \mu  ^\pm \nu \nu \nu $~\cite{Artamonov:2016wby} constraining the branching ratio to $ \lesssim   2.4 \times 10 ^{ - 6} $ (improving on a limit set in \cite{Pang:1989ut}). Since the neutrino and $ X $ channels share a similar final state, we use this limit to approximate our bound. We note that the constraints may be able to be improved by searching for the electron channel due to reduced Standard Model backgrounds. 

\begin{figure*} 
  \begin{center} 
\includegraphics[width=14cm]{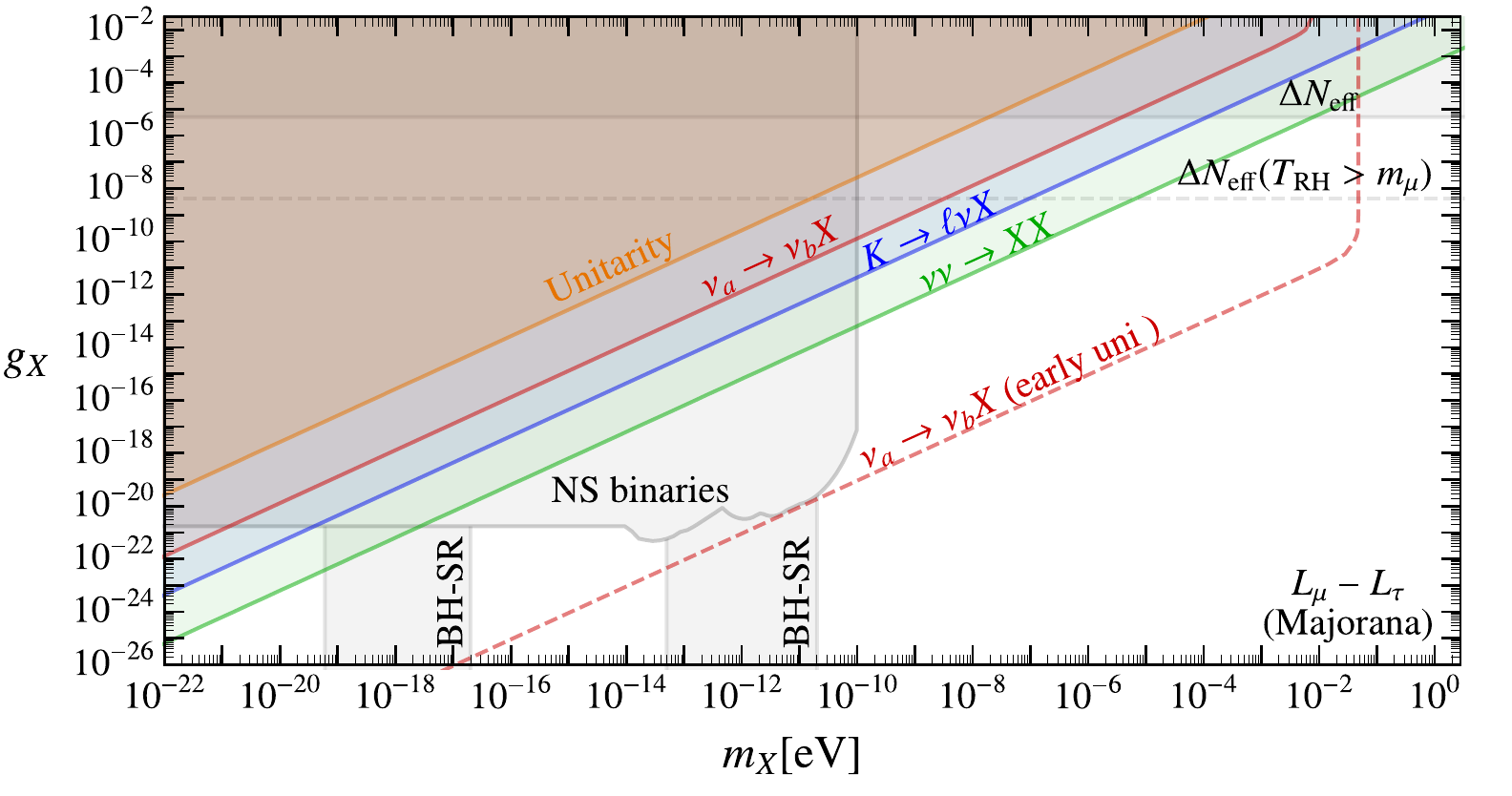}
\end{center}
\caption{Constraints on $ L _\mu  - L _\tau   $ for Majorana neutrinos. Previous constraints are shown in {\bf \color{cgray} gray} and are from neutron star binaries~\cite{Dror:2019uea}, black hole superradiance~\cite{Baryakhtar:2017ngi}, and $ \Delta N _{ {\rm eff}} $ measured through Big Bang Nucleosynthesis, with the latter depending on whether the universe reheated above the muon mass~\cite{Escudero:2019gzq} (dashed) or below~\cite{Huang:2017egl} (solid). Enhanced constraints come from unitarity ({\bf \color{corange} orange}) meson decays ({\bf \color{cblue} blue}) with $ K \rightarrow \ell \nu X $ being most stringent, $ \Delta N _{ {\rm eff}} $ measured with Big Bang Nucleosynthesis due to enhanced neutrino annihilations ({\bf \color{cgreen} green}), and neutrino decays ({\bf \color{cred} red}). The neutrino decay bounds arise from both terrestrial (solid) and cosmological (dashed) searches, the latter which assumes the neutrinos are present during recombination.}
\label{fig:LmuLtau}
\end{figure*}

The limits are depicted in the case of Majorana neutrinos for $ L _\mu  - L _\tau  $ and $ L _ e - L _\mu  $ in Figs. ~\ref{fig:LmuLtau} and ~\ref{fig:LeLmu}. Given their qualitative similarity we reserve the rest of the cases to appendix~\ref{app:additional}. We see that both these constraints are a powerful bound on the parameter space. Furthermore, they are insensitive to the UV completion, do not make any assumptions about cosmology, and apply to both Majorana and Dirac neutrinos. This robustness against the modifications of the model makes this channel important in searching for light $ L _i - L _j $ vectors.

\section{Neutrino decays ($ \nu _a \rightarrow \nu _b X $)}
\label{sec:nudecay}
Gauging $ L _i - L _j $ induces (neutrino mass-suppressed) lepton-number violating processes. The simplest possibilities are invisible neutrino decays, as depicted in Fig.~\ref{fig:allprocesses} ({\bf bottom-left}), where a heavier neutrino decays into a lighter neutrino, emitting off a $  L _i - L _j $ vector. In principle its possible to look for visible neutrino decays instead of invisible decays through the neutrino magnetic moment, $ \nu _a \rightarrow  X ( \nu _b ^\ast \rightarrow \nu _b \gamma ) $. While the bounds on such decays are much more stringent, with searches looking for spectral distortions in the Cosmic Microwave Background (CMB) constraining lifetimes reaching $  {\cal O} ( 10 ^{ 20} {\rm sec})  $~\cite{Aalberts:2018obr}, the insertion of a magnetic moment suppresses the rate relative to the invisible decay by a parametric factor of $ \sim ( m _\nu / m _W ) ^4 $ which is too large a suppression to be relevant for us here and we focus on invisible decays.

The $ \nu _a \rightarrow \nu _b X $ decay rate into the longitudinal mode is, 
\begin{align} 
\Gamma _{ a \rightarrow b X} & = \frac{1}{ 16 \pi m _a } \overline{ \left| {\cal M} \right| ^2 } \frac{ \lambda ^{1/2} ( m _a ^2 , m _b ^2 , m _X ) }{ m _a ^2 }\,,
\end{align} 
where $ \lambda ( \alpha    , \beta  , \gamma  ) \equiv \alpha  ^2 + \beta  ^2 + \gamma  ^2 - 2 ( \alpha \beta + \beta \gamma  + \alpha \gamma  ) $. The spin-averaged matrix element squared has a mild dependence on whether neutrinos are Majorana (top line) or Dirac (bottom line):
\begingroup
\renewcommand*{\arraystretch}{1.75}
\begin{align} 
\overline{ \left| {\cal M} \right| ^2 } & = \frac{ g _X ^2  }{ m _X ^2   } \left\{ \begin{array}{lr}  ( m _a ^2 - m _b ^2 ) ^2 \text{Re} Q _{ ab } ^2 + ( m _a + m  _b ) ^4 \text{Im} Q _{ ab } ^2    \\ \frac{1}{2} \left|  Q _{ ab}  \right| ^2 ( m _a ^2 - m _b ^2 ) ^2 \end{array} \right.\,,
\end{align} 
\endgroup
where $ \text{Re} Q _{ ab }  $ and $ \text{Im} Q _{ ab } $ denote the real and imaginary parts of $ Q _{ ab } $. In the limit the PMNS matrix is real, the rates differ by a factor of $ 2 $.

The strongest constraint on invisible neutrino decays come from cosmology. If neutrinos are present during recombination then their decay, in combination with subsequent coalescence with $ X $, can prevent free-streaming from efficiently wiping out structure at small enough-$ \ell $ in Cosmic Microwave Background~\cite{Hannestad:2004qu,Hannestad:2005ex}. A recent study using Planck 2018 data~\cite{Escudero:2019gfk} set a bound on the lifetime, 
\begin{equation} 
\Gamma _{ a \rightarrow b X } ^{-1} \gtrsim 1.3 \times 10 ^{ 9 } ~{\rm sec} ~\left( \frac{ m _a }{ 0.05~{\rm eV}} \right) ^3  \,.
\end{equation} 
These are the strongest limits on $ L _i - L _j $ through enhanced processes. In addition, the cosmological bound could be drastically improved with the observation of a nonzero neutrino mass sum~\cite{Serpico:2007pt,Chacko:2019nej,Chacko:2020hmh}. A null result would improve the robustness of the cosmological bound and resultantly significantly change the allowed parameter space making cosmological neutrino lifetime measurements the most promising avenue to discover ultra light $ L _i - L _j $ vectors.

While significant, the present constraints assume neutrinos are around during recombination, which need not be the case if additional decay or annihilation channels were active after Big Bang Nucleosynthesis producing other purely free-streaming sources of radiation~\cite{Beacom:2004yd}. Alternatively, one could consider cases where the neutrino masses vary in the late universe~\cite{Gu:2003er}, rendering them stable during recombination. While perverse, these options reflect a clear model-dependency in the constraints and hence we also include constraints on the neutrino lifetime from terrestrial experiments. The strongest such bound is from studying the flavor composition of solar neutrinos~\cite{Beacom:2002cb}. Observations of the Sudbury Neutrino Observatory in combination of other solar neutrino experiments~\cite{Aharmim:2018fme}, constrain the lifetime of the second neutrino mass eigenstate,
\begin{equation} 
\Gamma _{ \nu _2 \rightarrow \nu _1 X  } ^{-1} \gtrsim 1.04 \times 10 ^{ - 3} ~{\rm sec} \left( \frac{ m _2 }{ {\rm eV} } \right) \,,
\end{equation} 
while the constraint on $ \nu _3 $ decay is considerably weaker~\cite{Funcke:2019grs}. The huge gap between the ability of cosmological and terrestrial bounds will be reduced by roughly four orders of magnitude with future studies of the flavor content of astrophysical neutrinos at IceCube, if astrophysical uncertainties can be sufficiently well understood~\cite{Beacom:2002vi,Pagliaroli:2015rca,Bustamante:2016ciw} (with even some hints at decays in current data \cite{Denton:2018aml}).

\begin{figure*} 
  \begin{center} 
\includegraphics[width=14cm]{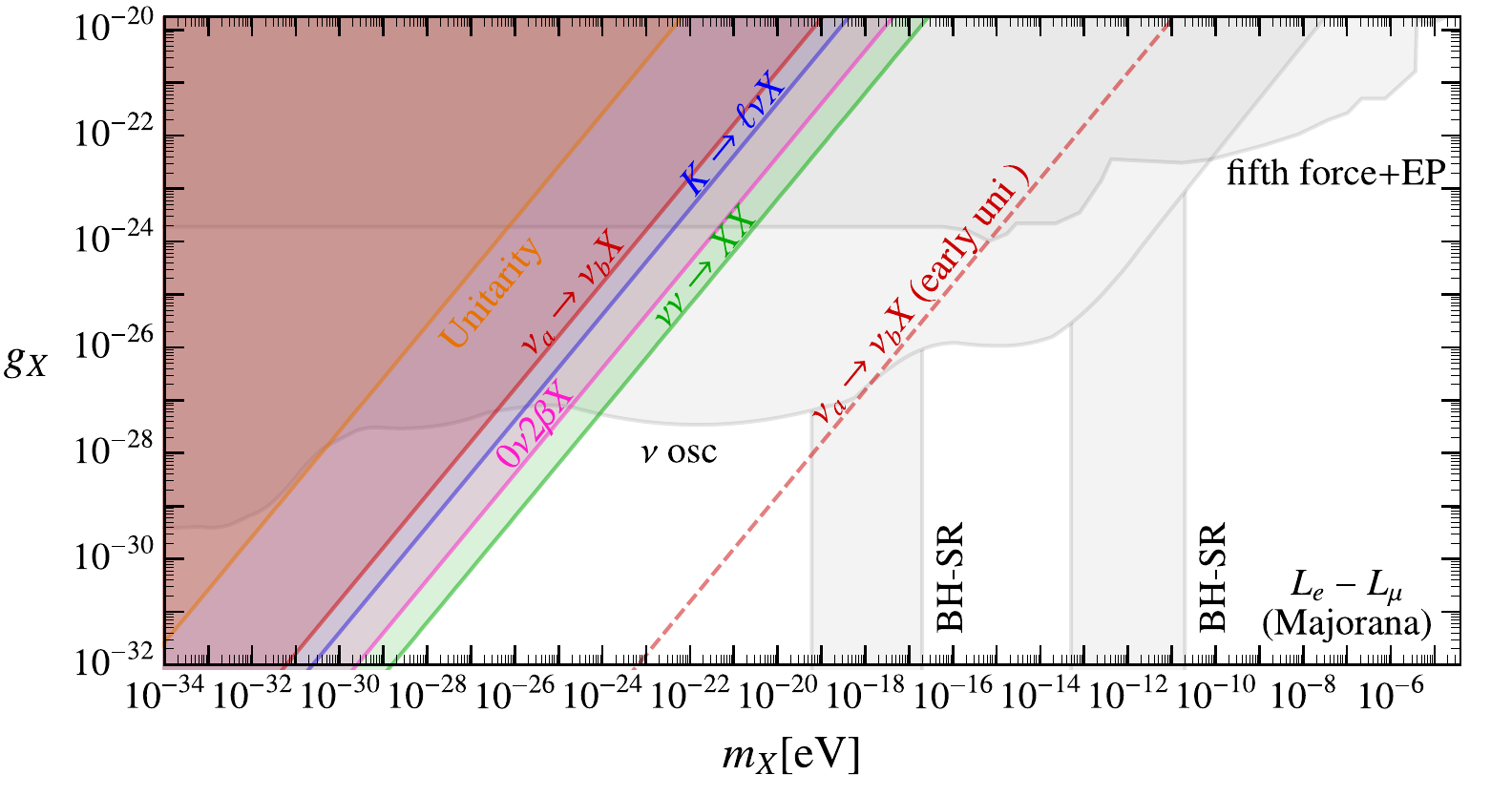}  
\end{center}
\caption{Constraints on $ L _e  - L _\mu    $ in the case of Majorana neutrinos. In {\bf \color{cgray} gray} are previous constraints from fifth forces~\cite{Wise:2018rnb}, deviations to neutrino oscillation data~\cite{Bustamante:2018mzu}, and black hole superradiance~\cite{Baryakhtar:2017ngi}. The enhanced constraints come from unitarity ({\bf \color{corange} orange}), meson decays ({\bf \color{cblue} blue}), dominated by $ K \rightarrow \ell \nu X $, neutrinoless double beta decay searches ({\bf \color{cpink} pink}), Big Bang Nucleosynthesis/supernova ({\bf \color{cgreen}green}), and neutrino decays ({\bf \color{cred} red}). Neutrino decay bounds depend on whether or not neutrinos are present during recombination (dashed vs solid)}
\label{fig:LeLmu}
\end{figure*}

We show the constraining power of these processes in red in Fig.~\ref{fig:LmuLtau} for $ L _\mu - L _\tau $ and in Fig.~\ref{fig:LeLmu} for $ L _e - L _\mu $ with the cosmological bound as a dashed line and the, more robust, terrestrial bound as a solid line. 

\section{$ 0 \nu 2 \beta X $ decays}
\label{sec:0nu2b}
If neutrinos are Majorana-type, then $ L _i - L _j $ vectors can induce in neutrinoless double $ \beta $ decays. For $ L _e - L _\mu $ and $ L _e - L _\tau $ the dominant production mode is single $ X $ emission as shown in Fig.~\ref{fig:allprocesses} ({\bf top-right}). For $L _\mu - L _\tau $ this process is forbidden. This is easiest to see when working in the flavor basis, where the $ W $ and $ X $ interactions are both flavor diagonal and all the mixing is in the mass terms. Since the $X$ interactions are diagonal, and $ X $ does not couple to electrons for $ L _\mu - L _\tau $, an additional neutrino mass insertion is needed to induce neutrinoless double beta process making it highly suppressed. As such, the dominant mode for $ L _\mu - L _\tau $ is double-$ X $ emission.

In either case, $ X $-induced neutrinoless double beta decays are only allowed for Majorana neutrinos, where the high energy limit matches precisely to that of a Majoron with appropriate identification of the couplings, allowing us to make use of dedicated Majoron studies. Searches for neutrinoless double beta decays in association with a Majoron have been carried out by KamLAND-Zen~\cite{Gando:2012pj} and NEMO~\cite{GarciaiTormo:2011et,Arnold:2013dha} with the strongest limit from the former, setting the bound on the Majoron coupling to electron neutrinos, $ \left| g _{ ee} \right| \lesssim (0.8 - 1.6 ) \times 10 ^{ - 5 } $, depending on the nuclear matrix element. The corresponding coupling in terms of $ L _i - L _j $ parameters is, 
\begin{equation} 
g _{ ee} \leftrightarrow  \frac{ g _X \hat{ Q} _{11} \hat{m} _{ 11} }{ m _X }
\end{equation} 
 (recall that the hats denote the matrices within the flavor basis). The prospective sensitivity to neutrinoless double beta decays with multiple-$ X $ emission was considered for Majorons~\cite{Bamert:1994hb} and is comparable in overall rate sensitivity, however, since the rate is proportional to $ \left( g _X m _\nu / m _X \right) ^4 $, we do not expect these to significantly improve the bounds given other constraints and as such we do not include them here.  

The sensitivity of $ 0 \nu 2 \beta X$ searches is shown in pink in Fig.~\ref{fig:LeLmu}. As discussed above, these do not apply for Dirac neutrinos, nor for $ L _\mu - L _\tau $.

\section{Neutrino annihilations} 
\label{sec:ann}
$ X $ bosons can also be produced through neutrino annihilations in a thermal bath in the early universe or during a supernova, as depicted in Fig.~\ref{fig:allprocesses} ({\bf bottom-right}). If both $ X $s are transverse then the amplitude is $ \propto g _X ^2  $, if one $ X $ is transverse and the other is longitudinal, then the amplitude is $ \propto g _X ^2  m _\nu / m _X  $ while if both $ X $s are longitudinal then the diagram is $ \propto g _X ^2 m _\nu ^2 / m _X  ^2 $. Since we focus on the case where $ g _X \ll 1 $, the dominant contribution will be from double-longitudinal emission and we neglect the rest.

Unlike the rest of the processes discussed so far, which can be computed in either the Goldstone boson equivalent Lagrangian or using the vector Lagrangian, double-$ X $ emission processes must be computed using the Goldstone boson picture to get a reasonable estimate of the rate (without providing a UV completion for the source of neutrino masses). The reason for this can be traced back to additional contribution from the UV completion to the divergence of the current in \eqref{eq:div1},\eqref{eq:div2}. In a U(1)$ _{ L _i - L _j } $-invariant model there must be a condensate coupling to $ X ^\mu $ (e.g., a $ L _i - L _j $-breaking elementary scalar). Since the condensate breaks the symmetry, it must also have linear couplings to $ X _\mu X ^\mu $ and will always contribute to processes with double-$ X $ emission. These are crucial since otherwise one may mistakenly conclude that double-longitudinal emission amplitudes for $ \nu \nu \rightarrow XX $ scattering can scale be $ \propto \, 1 / m _X  $,  as opposed to the expected $ \propto 1 / m _X ^2 $. This confusion does not arise if one works in the Goldstone boson equivalent theory since then the condensate coupling is not needed to ensure this cancellation. 

Note that a $ U(1)_{ L _i - L _j } $-breaking condensate will not have a significant impact on single-$ X $ emission considered above since single-$ X $ emission using additional scalars must arise from operators with derivatives on the scalar field(s) which can only interfere with the contributions we computed at loop level. Aside from this somewhat technical point this highlights an important point: bounds from neutrino annihilations have additional $ {\cal O} ( 1) $ uncertainties due to the presence of additional diagrams that contribution at the same order. In this sense multiple-$ X $ emission bounds are less robust, prior to choosing a UV completion, than single-$ X $ emission bounds. Keeping this in mind, we can still roughly estimate the bounds using places where neutrino scattering to invisible states can be actively measured, namely, Big Bang Nucleosynthesis and supernova.

Since we are interested in the high energy limit, we can again make use of the literature on Majorons and neutrino-philic scalars. Bounds from Big Bang Nucleosynthesis (BBN) on Majorons were computed in~\cite{Huang:2017egl}. Assuming flavor diagonal couplings they find the Majoron coupling to must be $ \lesssim 2 \times 10 ^{ - 5} $ at low masses to avoid disturbing the measured light element abundances. The constraints for $ L _i - L _j $ vectors will be similar with appropriate re-weighting of the coupling. To estimate the bounds we set this equal to $ \sim g _X m _3 / \sqrt{3} m _X $, where the factor of $ 1/ \sqrt{3} $ attempts to account for the fact that only one neutrino is interacting to first approximation (the most massive one). We require the same bound for Dirac neutrinos. The bounds are then shown in Fig.~\ref{fig:LmuLtau} and Fig.~\ref{fig:LeLmu} for $ L _\mu - L _\tau $ and $ L _e - L _\mu $ respectively.

 Supernova cooling bounds using supernova 1987A were computed for Majorons in the massless case~\cite{Farzan:2002wx}, constraining Majoron couplings to the level of $ {\cal O} ( 10 ^{ - 6} )  $. These could slightly strengthen the bounds from BBN, however are sensitive to the assumption on the inner temperature and could potentially be irrelevant if there was no proto-neutron star present after the supernova at all \cite{Bar:2019ifz}. Since these do not improve the bounds significantly, we do not include these but note that a supernova in our galaxy or improvements in supernova simulations would improve upon the annihilation bounds. 

There are a few other scattering processes one may consider. Charged lepton annihilations can proceed through $ \ell \ell \rightarrow XX $ and $ \ell \ell \rightarrow X \gamma $ however these will not be enhanced at tree level. In addition, neutrinos scattering off leptons or quarks through virtual electroweak boson exchange and radiating off a single $ X $ can result in enhanced processes. These will be efficient in a thermal bath at high enough temperatures, however, any limits set to this way will be inherently sensitive to the physics prior to BBN and will not be a robust bound on the parameter space. Alternatively, one could try to use neutrino-nucleus or neutrino-electron scattering experiments, however these experiments do not have the sensitivity necessary to probe the parameter space considered here.

\section{Discussion}
\label{sec:disc}
In this work we presented new constraints on $  L _i - L _j $ using the non-conserved nature of the currents. These bounds get stronger for smaller vector masses, making them particularly relevant for searches for long-range forces associated with gauging differences in lepton number. In particular, with current data we find that for $ L _e - L _\mu $ or $ L _e - L _\tau $, the new limits are now the strongest for $ m _X \lesssim 10 ^{ - 18 } {\rm eV} $ with a traditional cosmology or $ m _X \lesssim 10 ^{ - 24} {\rm eV} $ with a modified cosmology to evade CMB limits on the neutrino lifetime. For $ L _\mu - L _\tau $, where searches for long-range forces are less constraining, we find the new limits are the most important bounds on the parameter space for $ m _X \lesssim 10 ^{ - 2} {\rm eV} $ with a traditional cosmology or for a modified cosmology: $ 10 ^{ - 10} {\rm eV} \lesssim m _X \lesssim 10 ^{ - 2 }{\rm eV} $ in conjunction with $ m _X \lesssim 10 ^{ - 18} {\rm eV} $.

In either case our results have profound implications on the prospects of detecting $ L _i - L _j $ vectors using neutrino oscillation searches. For gauged $ L _e - L _\mu $ or $ L _e - L _\tau $, the bounds are no longer the strongest to arbitrarily small masses and observing such long-range forces on galactic scales~\cite{Bustamante:2018mzu} is not possible. Interestingly, if a measurement of the neutrino lifetime improves to cosmological scales (e.g., if a non-vanishing sum of neutrino mass is detected), the new bound could make it impossible to see deviations to neutrino oscillations even from the Earth/Sun. This would require a measurement of $ \tau _{ \nu _a \rightarrow \nu _b X } \gtrsim 10 ^{ 8}~ {\rm year} $, within the reach possible with the upcoming Euclid Satellite in combination with Planck data~\cite{Chacko:2020hmh}.

In addition, there have been proposals to probe gauged $ L _\mu - L _\tau $ with a mass-mixing of the vector with the $ Z $ boson (i.e., introducing a term $ \varepsilon _Z m _Z ^2 X _\mu Z ^\mu $), and again looking for deviations to neutrino oscillations patterns constraining the product, $ \varepsilon _Z g _X \lesssim {\cal O} ( 10 ^{ - 51} ) $ for $ m _X \lesssim 10 ^{ - 15} {\rm eV} $~\cite{Heeck:2010pg,Davoudiasl:2011sz}. Combining previously studied bounds on the mass-mixing parameter, $ \varepsilon _Z \lesssim 10 ^{ - 30} ( m _X / 10 ^{ - 15} {\rm eV} )  $~\cite{Dror:2018wfl} with CMB bounds on neutrino lifetime from this work, $ g _X \lesssim 10 ^{ - 24} ( m _X / 10 ^{ - 15} {\rm eV} ) $, we find searching for mass-mixed $ L _\mu - L _\tau $ is no longer viable. Relaxing the CMB bounds allows a small window for masses above the neutron star binary bounds.

We conclude by commenting on other prospective enhanced signals, which do not turn out to be effective at looking for $ L _i - L _j $ vectors. Firstly, one can look for lepton-number violating processes such as charged current decays at one-loop through neutrino mass insertions. For Dirac neutrinos, the chiral structure of the electroweak interactions forbid any contributions to the amplitude at order $ {\cal O} ( m _a ) $ leading to a suppression by an additional power of $ m _a / m _W $, rendering the decays unobservable. This suppression turns out to be present also for Majorana neutrinos. This is easiest to see in the Goldstone picture for $ X $. In this case, the only diagram that can contribute to $ \ell _a \rightarrow \ell _b X $ is from radiating off $ \varphi _X  $ from an internal neutrino leg. The chirality structure of the neutrinos and the weak interaction then requires an additional neutrino mass insertion for the process to take place. This suppression is significant enough to make this process uninteresting for our purposes. Alternatively, double-$ \varphi _X $ emission can take place however this is both phase space and double-neutrino mass suppressed making it a challenging way to discover $ L _i - L  _j $ forces. Lastly, one can look for enhanced $ W \rightarrow \ell \nu X $ decays as studied in \cite{Karshenboim:2014tka}. The $ W $ branching ratios are known to $ {\cal O} ( 0.1 \% ) $. When combined with the phase space penalty for producing a $ 3 $ body final state, this channel is not effective for looking for $ L _i - L _j $ vectors relative to other signals. 
\section*{Acknowledgments}

We thank Doug Bryman, Simon Knapen, Robert Lasenby, Toby Opferkuch, Maxim Pospelov, Nicholas Rodd, and Ofri Telem for useful discussions. This work was supported primarily by the DOE under contract DE-AC02-05CH11231. The study was initiated and performed in part at the Kavli Institute for Theoretical Physics, which is supported in part by the National Science Foundation under Grant No. NSF PHY-1748958. 
\appendix
\section{Goldstone boson equivalence}
\label{app:replacement}
In this section we prove that the following replacement rule for external $ X ^\mu $:
\begin{equation} 
{\cal L} _X \equiv   X _\mu j ^\mu \rightarrow   \frac{ g _X    \varphi _X }{ m _X } ( - i\partial _\mu j ^\mu )  \rightarrow  \frac{ g _X    \varphi _X }{ m _X } \hat{m} _{ ij } \hat{Q} _i \hat{\nu} _i \hat{\nu} _j \,.
\label{eq:replacement}
\end{equation} 
The switch from $ X ^\mu $ to its Goldstone mode $ \varphi _X $ follows from Goldstone boson equivalence theorem and we take it as a given. Our goal is then to prove that the divergence of the current can be replaced with the mass term, even when the fermions are off-shell. The derivation is in close analogy with that of the Schwinger-Dyson equations.

Consider a general correlation function, 
\begin{equation} 
{\cal Z}  =  \int {\cal D} \cdots (\psi  _1 \psi  _2 \cdots)  e ^{ i S ( X , \hat{ \nu}  , \hat{ \ell} ) }\,,
\end{equation} 
where $ S  $ contains the kinetic and mass terms for the fermionic and $ X $ fields as well as the source term $  \int \,d^4y j _\mu ( y ) X ^\mu ( y ) $. We use the shorthands, $ {\cal D} \cdots \equiv {\cal D} X {\cal D} \nu {\cal D} \ell $ and $ \psi _i = \hat{ \nu} _i  ( x _i )  $, $ \hat{ \ell} _i  ( x _i )  $, or $ X ( x _i ) $. Consider an infinitesimal field redefinition corresponding to a local U(1)$ _{L _i - L _j }  $ rotation,
\begin{align} 
\hat{ \nu} _i & \rightarrow (1+i \hat{Q}  _i \alpha (x) ) \hat{ \nu} _i \,, \\ 
\hat{ \ell}  _i & \rightarrow (1+i \hat{ Q} _i \alpha (x) ) \hat{ \ell}  _i 
\end{align} 
(leaving the vectors untouched). This field redefinition leaves the measure is unchanged since $ L _i - L _j $ transformations are anomaly free. Working in the flavor basis, its clear that the only terms in the action that are changed are the mass and interaction terms, i.e.,
\begin{equation} 
\delta S =    \int \,d^4y -  \partial _\mu \alpha (y)  j ^\mu  (y) -  i \hat{ m} _{ ij } \hat{ Q} _i \hat{ \nu} _i (y)  \hat{ \nu} _j  (y)  +{\rm h.c.}\,.
\end{equation}  
We take the neutrinos to be Majorana form here, though a similar equation will hold for Dirac neutrinos. 

In terms of these rotated fields, $ {\cal Z}  $ takes the form,
\begin{align} 
{\cal Z} & =  \int {\cal D} \cdots e ^{ i S + i \delta S }  ( 1 + \hat{ Q} _{ \psi _1 } \alpha _a ) \psi _1 ( 1 + \hat{ Q} _{ \psi _2 } \alpha _a )\psi _2 \cdots\,.
\end{align} 
Expanding the exponential and rearranging, 
\begin{align} 
 \delta {\cal Z} & = \int {\cal D} \cdots e ^{ iS } ( \psi _1 \psi _2\cdots ) \int \,d^4y    \bigg(  \partial _\mu j ^\mu - i \hat{ m} _{ ij } \hat{ Q} _i \hat{ \nu} _i \hat{ \nu} _j  \notag \\ 
& \qquad  +  ( \hat{ Q} _{ \psi _1 } \delta ( y - x _1 )  + \hat{ Q} _{ \psi _2 } \delta ( y - x _2 )  ) + \cdots \bigg) \alpha (y)\,.
\end{align} 
Since $ \alpha $ is arbitrary this should hold for any $ \alpha $ allowing us to drop the integral. Furthermore, since a correlation function should be invariant under a field redefinition, $ \delta {\cal Z} = 0 $, and we have, 
\begin{align} 
0 & = \int {\cal D} \cdots e ^{ iS } ( \psi _1 \psi _2 \cdots ) \bigg(      \partial _\mu j ^\mu (x) - i \hat{m}  _{ ij } \hat{ Q} _i \hat{ \nu} _i (x) \hat{ \nu} _j (x) \notag \\ & \qquad +  ( \hat{ Q} _{ \psi _1 } \delta ( x - x _1 )    + \hat{ Q} _{ \psi _2 } \delta ( x - x _2 )  ) + \cdots\bigg)\,.
\end{align} 
We conclude that, 
\begin{align} 
& - i \braket{ \Omega | T  \partial _\mu j ^\mu ( x ) \psi _1 \psi _2 \cdots  | \Omega } = \notag \\  
& \hspace{0.15cm} \hat{m}  _{ ij } \hat{ Q} _i \braket{ \Omega | T  \hat{ \nu} _i \hat{ \nu} _j ( x ) \psi _1 \psi _2 \cdots   | \Omega } +{\rm h.c.} \hspace{0.1cm} \left( +\begin{tabular}{c}\text{contact} \\ \text{terms} \end{tabular} \right) \,.
\end{align} 
This proves the replacement rule~\eqref{eq:replacement}. A similar derivation applies for the case of Dirac neutrinos, giving 
\begin{align} 
& - i \braket{ \Omega | T \left[ \partial _\mu j ^\mu ( x ) \psi _1 \psi _2 \cdots \right] | \Omega }  = \notag \\ 
& \hspace{0.15cm} ( \hat{m}  _{ ij } \hat{ Q} _j + \hat{m} _{ ij} \hat{ Q} _i ^c )  \braket{ \Omega | T \left[ \hat{  \nu} _i \hat{ \nu} _j ( x ) \psi _1 \psi _2 \cdots \right]  | \Omega }\,,
\end{align} 
in addition to contact terms.
\section{Additional results}
\label{app:additional}
\begin{figure*} 
  \begin{center} 
\includegraphics[width=8.5cm]{LeLmu_M.pdf}  \includegraphics[width=8.5cm]{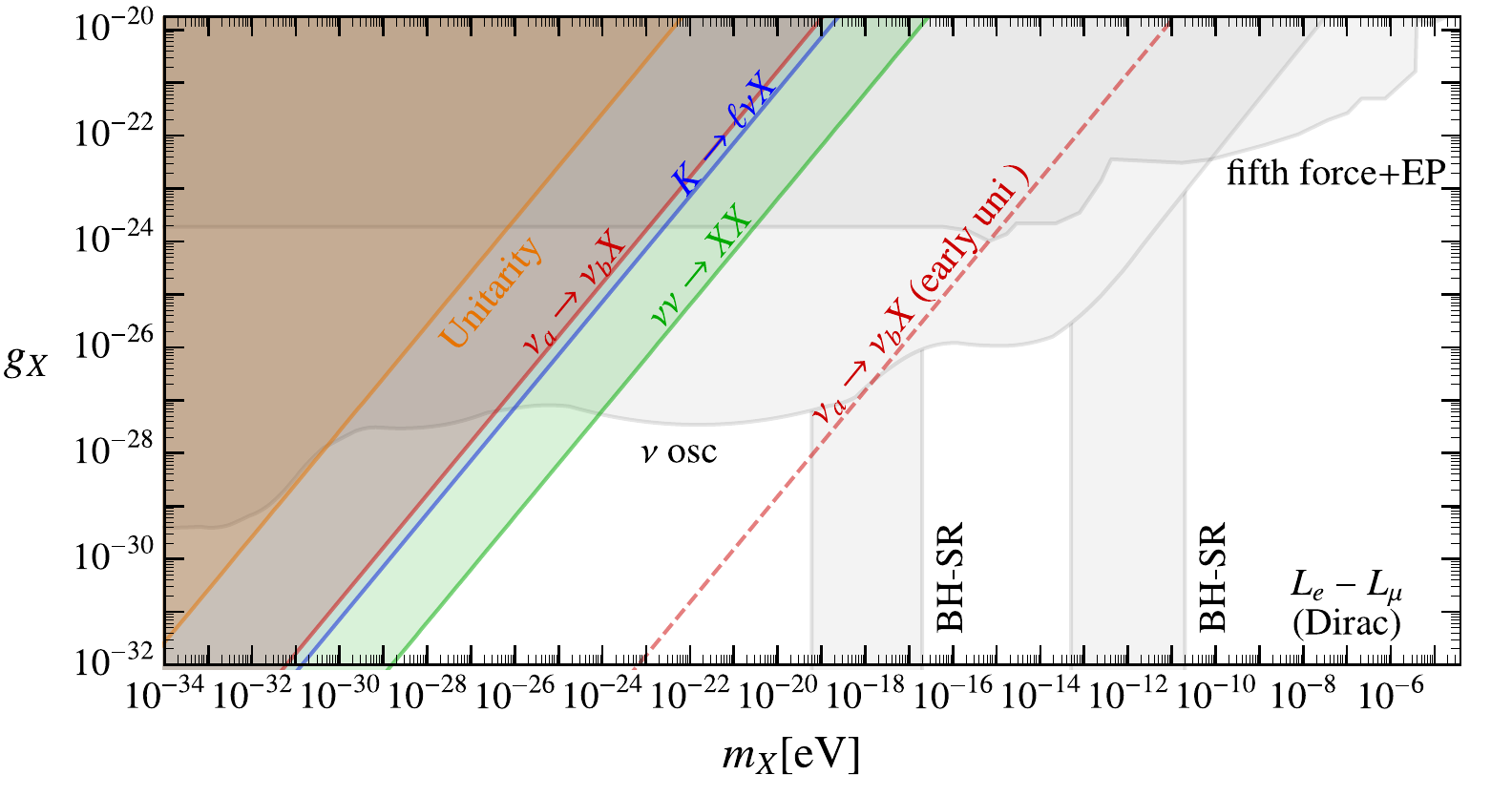}  
\includegraphics[width=8.5cm]{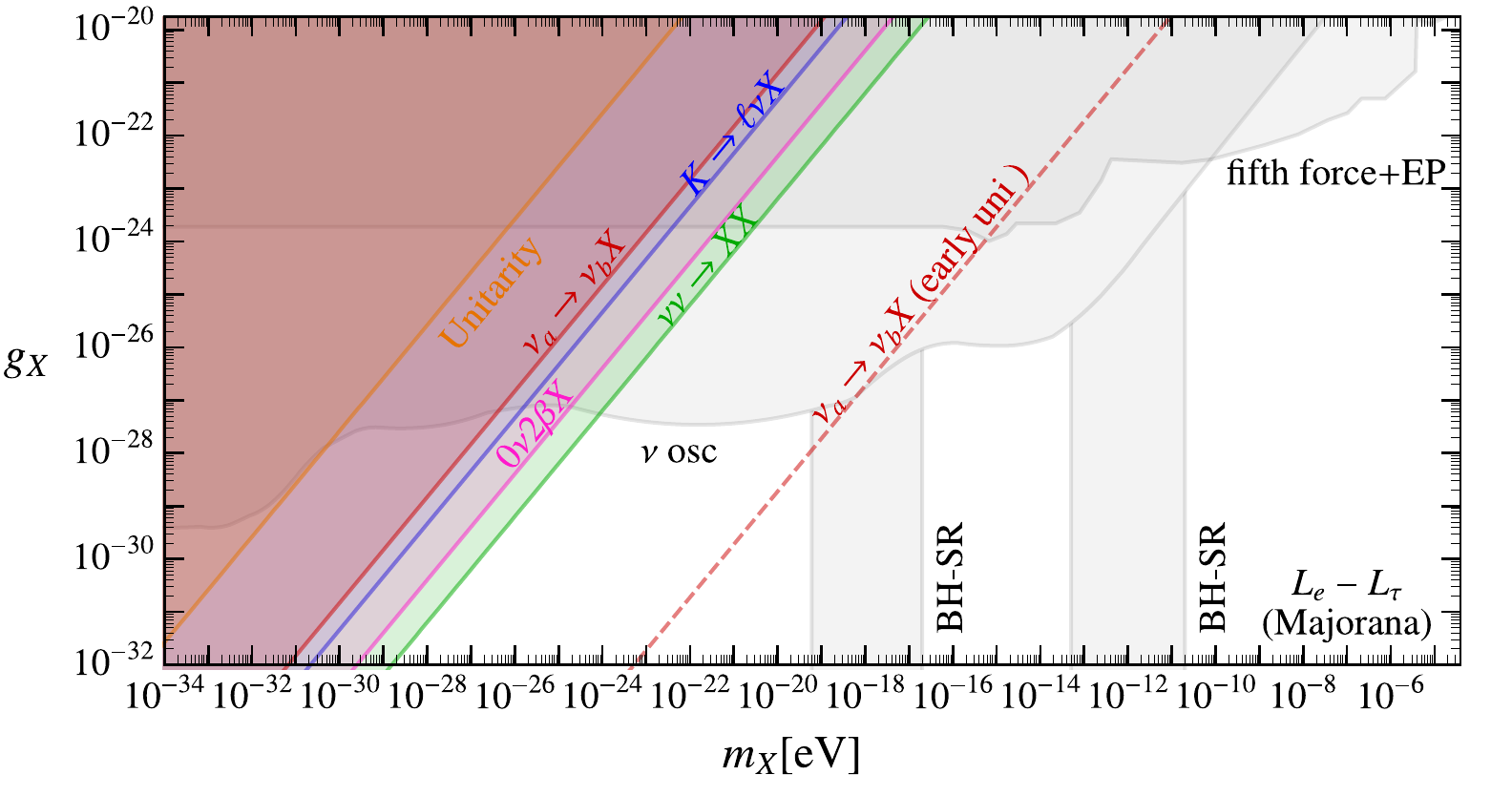}  \includegraphics[width=8.5cm]{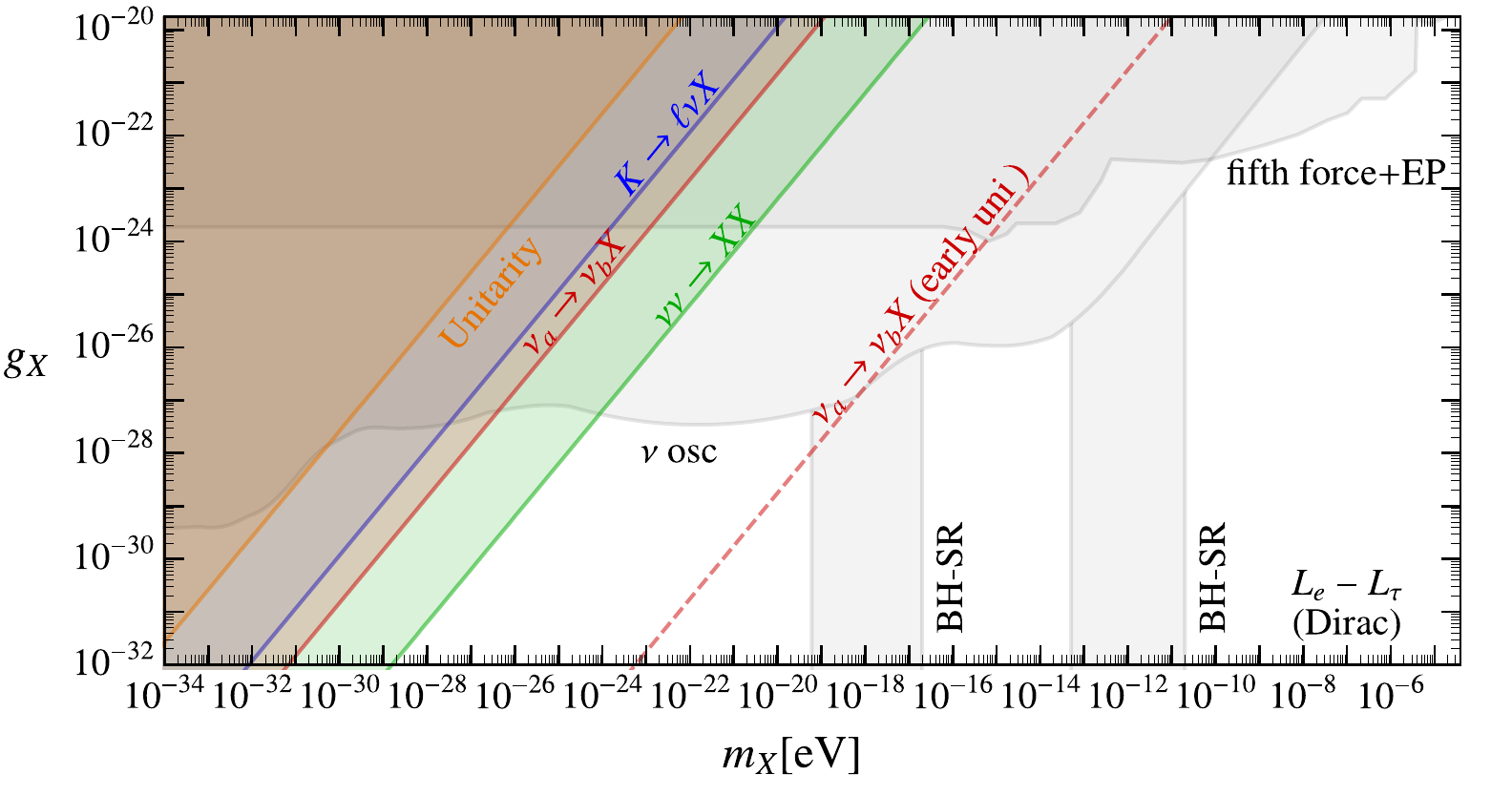}  
\includegraphics[width=8.5cm]{LmuLtau_M.pdf}  \includegraphics[width=8.5cm]{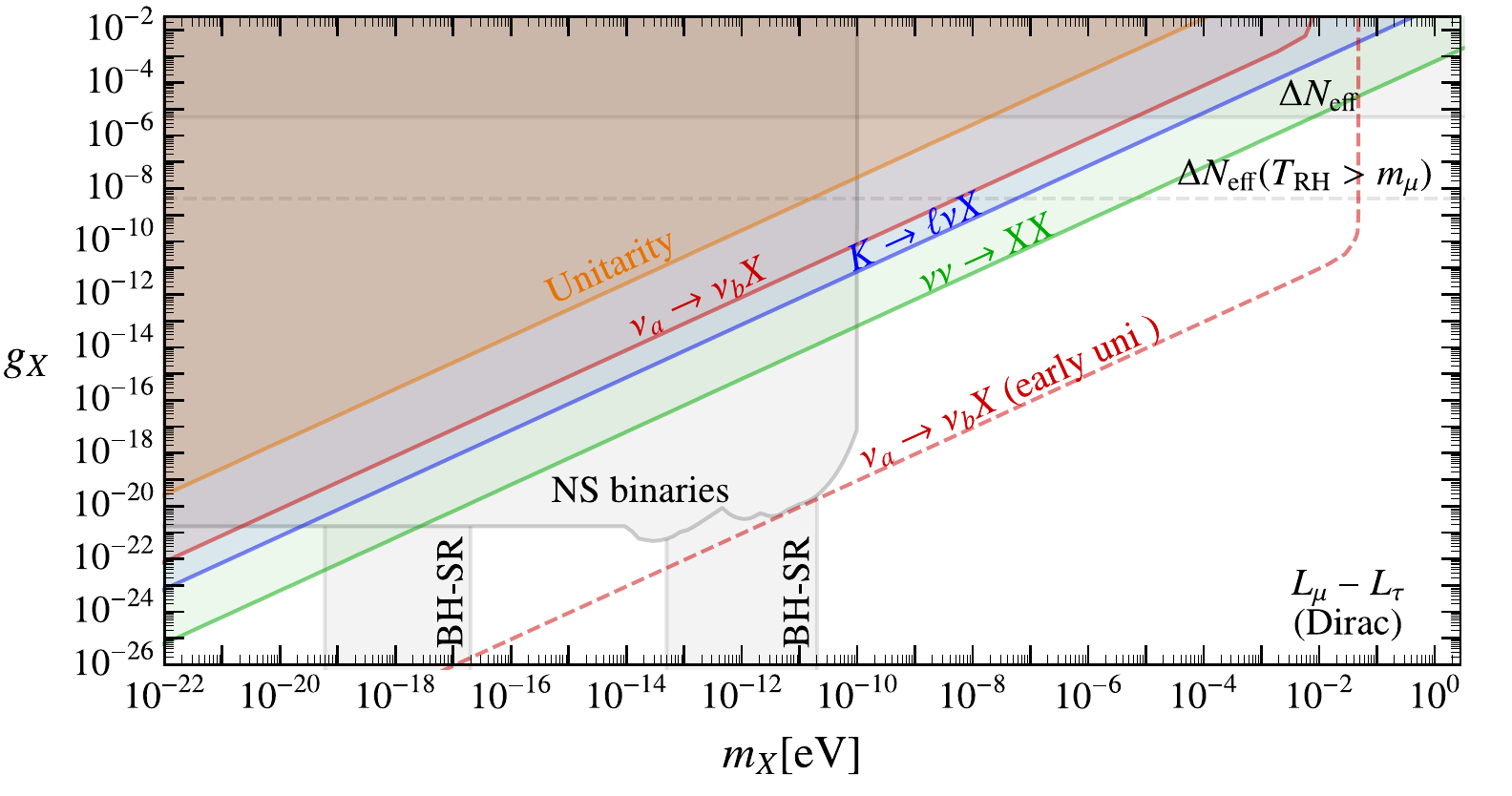}  
\end{center}
\caption{Constraints on $ L _i  - L _j     $ for Majorana and Dirac neutrinos. Previous constraints ({\bf \color{cgray} gray}) are from black hole superradiance~\cite{Baryakhtar:2017ngi}, neutron star binaries~\cite{Dror:2019uea}, fifth forces/Equivalence principle tests~\cite{Wise:2018rnb}, deviations to neutrino oscillations~\cite{Bustamante:2018mzu}, and $ \Delta N _{ {\rm eff}} $ measured through Big Bang Nucleosynthesis, with the latter depending on whether the universe reheated above the muon mass~\cite{Escudero:2019gzq} (dashed) or below~\cite{Huang:2017egl} (solid). The enhanced constraints come from unitarity ({\bf \color{corange} orange}), meson decays ({\bf \color{cblue} blue}), dominated by $ K \rightarrow \ell \nu X $, neutrinoless double beta decay searches ({\bf \color{cpink} pink}), Big Bang Nucleosynthesis/supernova ({\bf \color{green} green}), and neutrino decays ({\bf \color{cred} red}).}
\label{fig:all}
\end{figure*}

In the main text we only showed the constraints on $ L _e - L _\mu $ and $ L _\mu -  L _\tau $ and only for Majorana neutrinos. For those interested in quantitative limits we provide the limits for the other cases here. The summary of limits for $ L _e - L _\mu $, $ L _e - L _\tau $, and $ L _\mu - L _\tau $ are shown in Fig.~\ref{fig:all} in the {\bf top}, {\bf center}, and {\bf bottom} panels respectively. On the {\bf left} we present the limits for Majorana neutrinos and on the {\bf right} we present the limits for Dirac neutrinos. The non-enhanced bounds are all shown in gray and come a combination of fifth force/equivalence principle violation searches, cosmology, and long-range forces affecting neutrino oscillation patterns. In orange we show the unitarity bound, in blue the limits from meson decays (dominated by Kaon decays for all cases except $ L _e - L _\tau $ for Dirac neutrinos, where $ K \rightarrow \mu \nu X $ is suppressed by additional powers of neutrino masses), in pink neutrinoless double beta decay search limits, in green constraints from neutrino annihilations through observations of Big Bang Nucleosynthesis/supernova, and in red the bounds from neutrino decays (the cosmological bounds are dashed).

 \bibliographystyle{JHEP}
\bibliography{LiLjX}

\end{document}